\documentstyle[epsfig]{aipproc}
\def\Journal#1#2#3#4{{\em #1}{\bf #2}, #3 (#4)}

\def\PRD{Phys.~Rev. \bf{D}}

\begin{document}
\title{Re-analysis of $\pi\pi$/K$\pi$ Phase Shift
and Existence of
$\sigma$(600)/$\kappa$(900)-Particle}

\author{ 
Taku Ishida$^*$, Muneyuki Ishida$^{\dagger}$
Shin Ishida$^{\ddagger}$\\
Kunio Takamatsu$^{\star}$ and
Tsuneaki Tsuru$^*$}
\address{
$^*$
KEK, Oho, Tsukuba, Ibaraki 305, and
$^{\dagger}$
Department of Physics, University of Tokyo, Tokyo 113,\ 
$^{\ddagger}$
Atomic Energy Research Institite, College of Science and Technology,
Nihon University, Tokyo 101,\ and
$^{\star}$
Miyazaki U., Gakuen-Kibanadai, Miyazaki 889-21, JAPAN}

%\lefthead{LEFT head}
%\righthead{RIGHT head}
\maketitle

\begin{abstract}
Re-analyzing\cite{rf:pipip,rf:pik} 
the old phase shift data of I=0, $\pi\pi$
(I=1/2, $K\pi$) scattering, we show an evidence for existence 
of the $\sigma$-particle($\kappa$-particle) 
with a comparatively light mass, 
which has been missing for a long time:
In the analysis we have applied a new method of 
interfering Breit-Wigner amplitudes, which makes 
the scattering amplitude parametrized in terms of
only physical quantities, mass and width of resonant particles,
in conformity with unitarity. We have
introduced phenomenologically a background phase shift $\delta_{BG}$ 
of repulsive core type.
The $\chi^2$-fits of the both $\pi\pi$- and $K\pi$- 
phase shifts are improved greatly with the
$\delta_{BG}$ over the conventional fits without it.
\end{abstract}

\begin{figure}[t]
\epsfysize=8. cm
\centerline{\epsffile{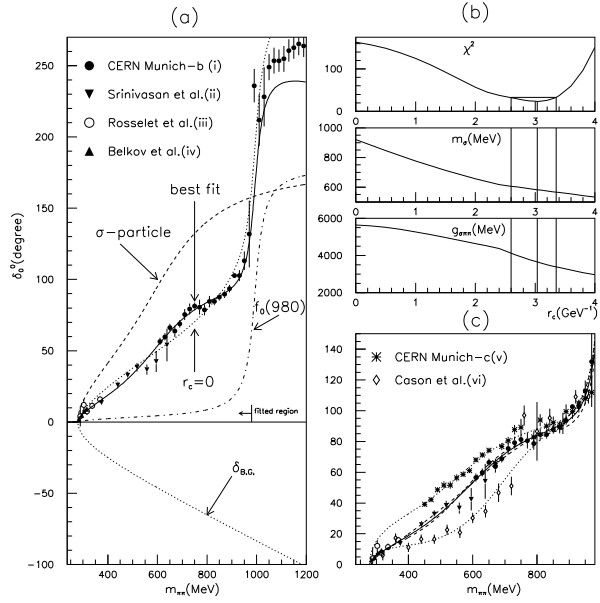}}
\caption{$I$=0 $\pi\pi$ scattering phase shift.
(a) Best fit to the standard $\delta_0^0$.
The dotted line labeled
``$r_c$=0'' represents the conventional fit without 
the repulsive background.
(b) $\chi^2, M_\sigma$ and $g_\sigma$ versus $r_c$. 
(c) Fits to the upper and lower $\delta_0^0$.}
\label{fig:i0}
\end{figure}
 The analyses are made by applying the new method
of $S$-matrix parametrization, IA-method:
The partial $S$-wave
$S$ matrix element in the $\pi\pi$ system 
is represented by the phase shift $\delta (s)$
and the amplitude $a(s)$ or ${\cal T}(s)$, as
\begin{eqnarray}
S &=& e^{2i\delta (s)} = 1+2ia(s)= 1+2i\rho {\cal T}(s)\ 
(\rho_1\equiv p_1/8\pi\sqrt{s},\ \ p_1\equiv\sqrt{s/4-m_\pi^2}). 
\label{eq:ddef}
\end{eqnarray}
The $\delta$ is given by the sum of
$\delta_{\rm Res}$ and $\delta_{\rm BG}$, respectively, due to
each resonance and the background.
Correspondingly, the total ${\cal S}$ matrix is
given by the product of 
individual ${\cal S}$ matrices. 
\begin{eqnarray} 
\delta = \delta_{\rm Res}+\delta_{\rm BG},\ 
\delta_{\rm Res} = \stackrel{(\sigma)}{\delta}+
\stackrel{(f_0)}{\delta}. \ \
S = S^{\rm Res}S^{\rm BG} = 
\stackrel{(\sigma)}{S}\stackrel{(f_0)}{S}S^{\rm BG}.
\label{eq:tots}
\end{eqnarray}
The unitarity of $S$ is
satisfied by the unitarity of individual
$S$ matrices.
Each $S^{\rm Res}$ is expressed by a relativistic 
Breit-Wigner (BW) amplitude,
\begin{eqnarray}
\stackrel{(R)}{a}(s) &=&
\frac{-\sqrt{s}\Gamma_R (s)}
{(s-M_R^2) + i\sqrt{s}\Gamma_R (s)}, \ \ \ 
\sqrt{s}\Gamma_R (s)\equiv \rho_1 g_R^2,
\label{eq:bwb}
\end{eqnarray}
\begin{table}[b]
\caption{Obtained parameter values and their errors.
$\Gamma^{(d)}$: decay width.
}
\begin{center}
\begin{tabular}{l|c|c|cc}
\hline
 & \multicolumn{2}{c|}
{standard $\delta_0^0$ (i)--(iv)} & upper bound (v) & lower bound (vi)\\
\hline
$M_\sigma$     &   585$\pm$20 MeV  & (920)  & 540 & 675 \\
$g_{\sigma\pi\pi}^r$  & 3600$\pm$350 MeV & (5650) & 3720 & 3570\\  
$\Gamma_{\sigma\pi\pi}^{(p)}$ & 385$\pm$70 MeV & (660) & 440 & 345\\
$\Gamma_{\sigma\pi\pi}^{(d)}$ &  340$\pm$45 MeV & (630) & 380  &  325 \\
$\sqrt{s_{\rm pole}}$ & (0.602$\pm$0.026)      & (0.97$-i$0.32) &- &- \\
         ~          & $-i$(0.196$\pm$0.027)  & & & \\
$r_c$    &   3.03$\pm$0.35 GeV$^{-1}$ & (0.) & 2.99 & 2.69 \\ 
 &0.60$\pm$0.07 fm & ~ & 0.59 & 0.53 \\
$\chi^2/N_{\rm d.o.f.}$ & 23.6/(34$-$4) & 163/(34$-$3) & 32.3/(26$-$4) & 42.1/(17$-$4) \\
\hline
\end{tabular}
\label{tab:mw}
\end{center}
\end{table}
where $\Gamma_R (s=M_R^2)$ represents the peak width
$\Gamma_R^{(p)}$ of the resonance $R$,
$g_R$ is the $\pi\pi$-coupling constant.
The amplitude $a^{\rm Res}$ (or ${\cal T}^{\rm Res}$)
is represented  by 
\begin{eqnarray}
a^{\rm Res} &=& \stackrel{(\sigma )}{a}+\stackrel{(f_0)}{a}
+2i\stackrel{(\sigma )}{a}\stackrel{(f_0)}{a}\ \ 
({\cal T}^{\rm Res} = \stackrel{(\sigma )}{{\cal T}}
+\stackrel{(f_0)}{{\cal T}}
+2i\stackrel{(\sigma )}{{\cal T}}\stackrel{(f_0)}{{\cal T}}),
\label{eq:IA}
\end{eqnarray}
which is the sum of the respective Breit-Wigner amplitudes
and their ``interfering" crossing term.
In the analyses, the $\delta_{\rm BG}$ is assumed to be
$\delta_{\rm BG}(s) = -|{\bf p}_1|r_c $
of repulsive core type.

The data of ``standard phase shift''
$\delta^0_0$ between $\pi\pi$- and K$\overline{\rm K}$-thresholds were used.
We also analyzed the data on upper and lower bounds
reported so far (see ref.\cite{rf:pipip} in detail).
Fig.~1(a) displays the results of the best fit
to the standard $\delta_0^0$, which includes the sum of three contributions, 
$\sigma$, $f_0(980)$, and $\delta_{\rm BG}$.

\begin{figure}[t]
 \epsfysize=8. cm
 \centerline{\epsffile{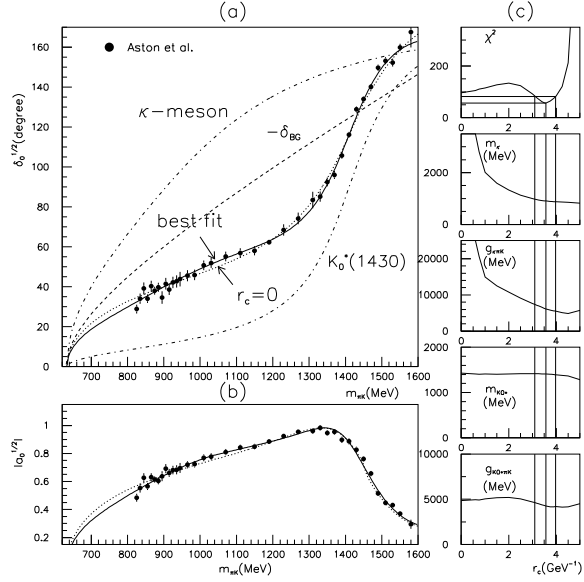}}
\caption{
Fits to {\it I}=1/2 $K\pi$ {\it S}-wave scattering amplitude; 
(a) phase shift $\delta_0^{{1/2}}$, and 
(b) magnitude of amplitude $|a_0^{1/2}|$.
(c) $\chi^2$, $M_\kappa$, $g_\kappa$, $M_{K_0^*}$, and $g_{K_0^*}$ behavior
as functions of core radius $r_c$.}
\label{fig:i12}
\end{figure}
\begin{table}[b]
\caption{Resonance parameters of $\kappa$(900), $K_0^*(1430)$ 
and core radius.
The errors correspond to five standard deviations.
}
\begin{center}
\begin{tabular}{l|ccccc}
\hline
 & $M_\kappa$/MeV  & $g_{K\pi}$/MeV  &
$\Gamma_{K\pi}^{(p)}$/MeV &
$\Gamma_{K\pi}^{(d)}$/MeV  &   $\sqrt{s_{\rm pole}}/$GeV \\
\hline
$\kappa$(900) & 905$\stackrel{\scriptstyle +65}{\scriptstyle -30}$   
              & 6150$\stackrel{\scriptstyle +1200}{\scriptstyle -650}$  
              & 545$\stackrel{\scriptstyle +235}{\scriptstyle -110}$   
              & 470$\stackrel{\scriptstyle +185}{\scriptstyle -90}$   
              & (0.875$\pm$0.075)$-i$(0.335$\pm$0.110) \\ 
$K_0^*$(1430) & 1410$\stackrel{\scriptstyle +10}{\scriptstyle -15}$   
              & 4250$\stackrel{\scriptstyle +380}{\scriptstyle -70}$  
              & 220$\stackrel{\scriptstyle +40}{\scriptstyle - 5}$   
              & 220$\stackrel{\scriptstyle +40}{\scriptstyle - 5}$    
              & (1.410$\pm$0.015)$-i$(0.110$\pm$0.010) \\
\hline
\multicolumn{6}{c}{
$r_{c0}^{{1/2}}$=3.57
$\stackrel{\scriptstyle -0.45}{\scriptstyle +0.40}$ GeV$^{-1}$ 
(0.70$\stackrel{\scriptstyle -0.09}{\scriptstyle +0.08}$ fm)
\ \ $\chi^2/N_{\rm d.o.f.}$ 57.0/(51$-$9)}\\
\hline
\end{tabular}
\label{tab:kp}
\end{center}
\end{table}
In Fig.~1(b) $\chi^2$, $M_\sigma$ and
$g_\sigma$ are plotted as functions of $r_c$.
The $\chi^2$ has a deep parabolic shape, and makes its minimum at 
3.03 GeV$^{-1}$, being close to 
the ``structural size"(charge radius) of pion$\sim$0.7 fm.
In the first row of Table~\ref{tab:mw} 
the parameter errors are quoted as 3-sigma deviation in
this $\chi^2$ behavior.
The results of the analysis on upper/lower bound data 
are shown in Fig.~1(c). From all of these studies, 
we may conclude that 
$M_\sigma$ is in the range of 535--675 MeV and
$r_c$ is between 2.7--3.0 GeV$^{-1}$.
Note that the fit with $r_c$=0 corresponds to
the conventional analyses without the repulsive
$\delta_{\rm BG}$ thus far made.
In this case the mass and width of ``$\sigma$"
becomes large, and the ``$\sigma$"-Breit-Wigner
formula can be regarded as an  
effective range formula 
describing a positive background phase.
The obtained values of parameters are  
$M_\sigma$=920 MeV and $g_\sigma$=5650 MeV
corresponding to the pole position
$\sqrt{s_{\rm pole}}$=(0.97$-$0.32$i$) GeV on sheet {\em II},
which is close to that of $\epsilon$(900), (0.91$-$0.35$i$) GeV,
in Ref.~\cite{rf:MP}. In this case the bump-like structure 
around $500\sim 600$ MeV in $\delta_0^0$ 
cannot be reproduced, as shown in 
Fig.~1(a), and gives $\chi^2$ = 163.4, worse by 140 than the best fit.
This seems to suggest the $\sigma$-existence strongly.

Taking the $SU(3)$ flavor symmetry into account, 
it is natural to expect the existence of a 
scalar meson nonet\cite{rf:sca}.
We also analyze the {\em I}=1/2 $K\pi$-scattering 
phase shift from a similar 
method to the $\pi\pi$ system.
The results are given in Fig.~\ref{fig:i12}:
We can also identify 
a low-mass resonance $\kappa$ with 
a mass of 900 MeV in the slowly increasing phases 
between the threshold and 1300 MeV, owing 
to the role of $\delta_{\rm BG}$,
whose absolute value is shown by a dotted line in Fig.~2(a).

\begin{figure}[t]
 \epsfysize=4. cm
 \centerline{\epsffile{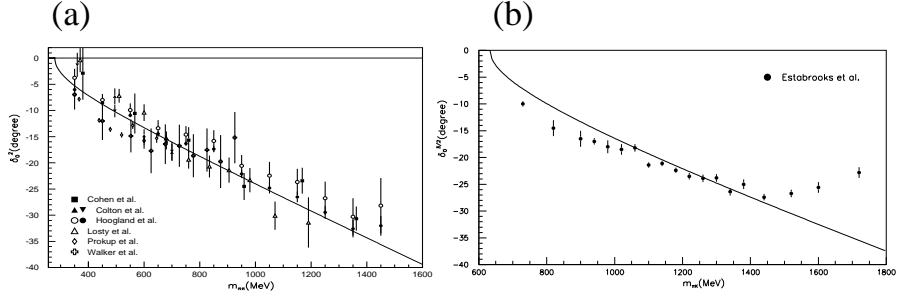}}
\caption{(a)$I$=2 and
(b)$I$ = 3/2 scattering phase shift.}
\label{fig:bg}
\end{figure}
In the present analyses
the introduction of a negative
background phase $\delta^{{\rm BG}}$ of 
the hard-core type plays an essential role.
In the $I$=2(3/2) system, there is
no known / expected resonance, and it is expected that
the repulsive core will appear directly.
In Fig.~\ref{fig:bg} the experimental
data of the 
$S$-wave phase shift for the $I$=2(3/2) scattering 
are shown, which are reproduced well
by the hard-core formula (solid lines),
with core radius of $r^2$=0.17fm ($r^{3/2}$=0.16fm).
It is interesting that they are almost same within
the non-exotic channels ($r^0$$\sim$$r^{{1/2}}$) and
within the exotic channels ($r^2$$\sim$$r^{3/2}$), respectively,
which seems to be reasonable from the viewpoint 
of $SU(3)$ flavor symmetry.
Historically this type of repulsive  $\delta_{BG}$ was found
in the cases of $\alpha$-nucleus-$\alpha$-nucleus scattering
and of nucleon-nucleon scattering.
In the relevant case of $\pi\pi$/K$\pi$ system the origin of $\delta_{BG}$
seems to have a strong connection to the ``compensating" 
$\lambda\phi^4$ interaction in the linear $\sigma$-model, required from
the current algebra and PCAC\cite{rf:pik,rf:MY}.

\end{document}